\documentclass[aps,prl,reprint, longbibliography]{revtex4-1}
\usepackage{color, soul}
\usepackage[latin9]{inputenc}
\usepackage[english]{babel}
\usepackage{float}
\usepackage{graphicx}
\usepackage{epstopdf}
\usepackage{amsmath}
\usepackage{esint}
\usepackage{natbib}
\makeatletter
\renewcommand{\fnum@figure}{FIG. \thefigure}
\makeatother

\begin{document}

\title{Mapping the landscape of domain-wall pinning in ferromagnetic films using differential magneto-optical microscopy}

\date{\today}

\author{Robert Badea}
\author{Jesse Berezovsky}
\email{jab298@case.edu}
\affiliation{Department of Physics, Case Western Reserve University, Cleveland, OH}

\begin{abstract}

The propagation of domain walls in a ferromagnetic film is largely determined by domain wall pinning at defects in the material. In this letter we map the effective potential landscape for domain wall pinning in Permalloy films by raster scanning a single ferromagnetic vortex and monitoring the hysteretic vortex displacement vs. applied magnetic field. The measurement is carried out using a differential magneto-optical microscopy technique which yields spatial sensitivity $\sim 10$~nm. We present a simple algorithm for extracting an effective pinning potential from the measurement of vortex displacement vs. applied field.  The resulting maps of the pinning potential reveal distinct types of pinning sites, which we attribute to quasi-zero-, one-, and two-dimensional defects in the Permalloy film.

\end{abstract}

\maketitle

\hrulefill

Studying the pinning of domain walls by defects in ferromagnetic films is critical for understanding and optimizing domain wall motion and magnetization switching. Pinning occurs when the domain wall, which is a quasi-linear region of high magnetostatic, exchange, and/or anisotropy energy density, encounters a defect which gives rise to a local minimum of the total energy\cite{Barkhausen1916, Hubert1998,Durin2006}. Here we study a simpler situation -- the pinning of a ferromagnetic vortex core. The ground state configuration of a thin, micron-scale ferromagnetic disk with negligible magnetocrystalline anisotropy is a vortex. This state is characterized by a curl of in-plane magnetization circulating about a central  core\cite{Cowburn1999, Shinjo2000, Wachowiak2002}. The vortex core, with half-width $\sim 10$~nm and magnetization oriented largely out of plane \cite{Hollinger2003}, is a point-like region of high energy density. As it is scanned across the sample via an applied magnetic field, the vortex core acts like a nanoscale probe for imaging the pinning potential\cite{Wachowiak2002, Burgess2013}.  The pinning effects studied here in ferromagnetic disks are likely similar to those in vortex domain walls in ferromagnetic nanowires\cite{Min2010}, with proposed applications in spintronic memory\cite{Parkin2008}. More generally, the vortex pinning maps that we obtain give insight into how pinning of extended domain walls would arise in these films, as the effects rely on similar energetic considerations. 

The interaction between a vortex state and a pinning site has been characterized for both strong, artificially created pinning sites\cite{Rahm2003, Rahm2004, Rahm2004b,Uhlig2005,Kuepper2007}, and weaker naturally-occurring pinning sites~\cite{Burgess2013,Chen2012, Chen2012b, Compton2006, Compton2010}. It was demonstrated that single point defects were capable of trapping the vortex core and thus suppressing its motion in response to applied magnetic field. While some information about the pinning potential can be extracted from these measurements, the techniques for data collection and analysis have generally been too cumbersome to extract the type of pinning potential maps presented here.

For small displacements, the motion of the vortex core in an in-plane magnetic field $\mathbf{H}$ can be described by the rigid vortex model\cite{Guslienko2001}, via a potential $U_0(\mathbf{r},\mathbf{H})\approx \frac{1}{2} k | \mathbf{r}|^2-k\chi_0(H_y x + H_x y) $, where $\mathbf{r}=(x,y)$ is the in-plane displacement of the core from the center, $k$ is the stiffness of the vortex which depends on the demagnetization factor of the disk, and $\chi_0$ is the displacement susceptibility of the free vortex. This potential describes a paraboloid with the position of the minimum shifted perpendicular to an applied in-plane magnetic field. 

\begin{figure}[htb]
\centering{}\includegraphics{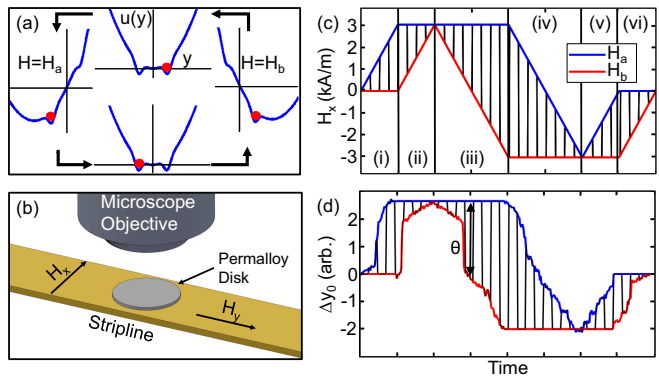}\protect\caption{{\footnotesize{}Measuring hysteretic vortex displacement. (a) Illustration of an effective 1D potential $u(y)$ with two pinning sites cycling between two fields $H_a$ and $H_b$. The red dot indicates the vortex position. (b) Diagram of the experimental geometry (not to scale).  (c) Sweep of AC magnetic field $H_x$ used to measure vortex displacement, with upper end ($H_a$) and lower end ($H_b$) of the 15-kHz square wave shown in blue and red respectively. Total measurement time $\sim 3$~min. (d) Vortex displacement $\Delta y_0$ in response to $H_x$ shown in (a). }} 
\label{cycling}
\end{figure}

Vortex pinning can be included in the model by adding localized potential minima $U_{p,i}(\mathbf{r})$ \cite{Burgess2014, Chen2012}. As a magnetic field $H_x$ is increased, the vortex core moves roughly in a line in the $y$-direction, with deviations from the line caused by the particular arrangement of pinning sites. It is convenient to represent the potential along this vortex path as an effective 1D potential $u(y,H_x)=u_0(y, H_x)+\sum{u_{p,i}(y)}$, with $u_0=\frac{1}{2} ky^2-k\chi_0 H_xy$.   For example, Fig.~\ref{cycling}(a) shows a possible effective 1D potential $u$ vs. $y$ cycling between two values of $H_x$, with two Gaussian pinning sites.  At temperature $T=0$, the vortex core resides at $y_0$, a local minimum of $u$, translating along with that minimum as $H_x$ is changed. The rate of displacement with applied field $\chi = dy_0/dH_x$ depends inversely on the curvature $u''$ at the local minimum. By definition, $\chi$ is suppressed when the vortex is pinned as compared to the unpinned case where $\chi = \chi_0$. 

Sudden changes in the pinning configuration occur with a change in $H_x$ when the presently occupied local minimum disappears (as illustrated in Fig.~\ref{cycling}(a)). Assuming the pinning sites are sufficiently spaced to be treated independently, the vortex leaves the $i$th pinning site when the maximum or minimum slope due to $u_{p,i}$ is cancelled by the slope of $u_0$, which occurs at pairs of values $(y^\pm,H_x^\pm)$ that satisfy 

\begin{equation}
u'(y^\pm,H_x^\pm)=u''(y^\pm)=0.
 \label{depinning}
\end{equation}

The $\pm$ indicates the two solutions with $u'_{p,i}(y^+,H_x^+)>0$ and $u'_{p,i}(y^-,H_x^-)<0$. As $H_x$ is swept through such a point, a sudden jump in $y_0$ and/or $\chi$ is observed. As illustrated in Fig.~\ref{cycling}(a), multiple local minima may be present at a particular value of $H_x$, resulting in hysteresis in $y_0$ as $H_x$ is swept up and down. At $T>0$, thermal activation of the vortex out of a local minimum is possible \cite{Eltschka2010}, causing smearing of the sudden jumps and a shift in the critical field at which the jumps occur, reducing the observed hysteresis \cite{Burgess2013}.

The samples measured here were Permalloy (Py, $\mathrm{Ni_{0.81}Fe_{0.19}}$) disks with diameter $d=2~\mu$m and thickness $t = 40$~nm.  Data from two samples, D1 and D2, are presented. Samples were were fabricated via electron-beam lithography, electron-beam evaporation, and liftoff atop a gold microstrip transmission line. The Permalloy was evaporated at a base pressure $P=2 \times 10^{-7}$~Torr and a rate $\tau_{d} = 0.05$~nm$/$s. The defects in the material resulting from the fabrication process act as the pinning sites for the vortex core. These include variations in the film thickness believed to be caused by residual contamination of the Au surface prior to Py deposition, smaller scale surface roughness of the film, and grain boundaries. We note that this technique could be applied more generally to samples on other substrates, which are then brought into close proximity with a stripline.

The experimental setup, with geometry shown in Fig.~\ref{cycling}(b), is based on a standard, room temperature magneto-optical Kerr effect (MOKE) microscopy apparatus \cite{Freeman2002}, described in more detail in Ref.~\cite{Badea2015}. The measurement is carried out in the longitudinal MOKE geometry via suitable masking of the probe laser which is focused through a high numerical aperture objective, and is therefore mainly sensitive to the magnetization component $M_x$.  With the probe spot focused on the vortex core position, $M_x=0$ by symmetry. As an in-plane field $H_x$ is applied, a nonzero $M_x$ appears which is roughly proportional to the vortex core displacement $\Delta y_0 = y_0(H_x)-y_0(H_x=0)$ for small displacement. 

In order to observe the hysteretic features in the vortex motion arising from the smallest pinning sites, it is necessary to achieve sufficient signal-to-noise to resolve changes in vortex position $\sim 10$~nm, the half-width of the vortex core and resolution limit of this technique. It is also necessary to carry out the measurement on sufficiently fast timescales to minimize thermal activation out of pinning sites. Both of these are accomplished using the differential nature of the experiment.  A 15-kHz square wave current is driven through the stripline beneath the magnetic disks, producing a magnetic field $H_x$ alternating between two values $H_a$ and $H_b$.  Using a lock-in technique, the measurement produces a signal $\theta \propto \Delta M_x = M_x(H_a)-M_x(H_b) \propto \Delta y_0 = y_0(H_a)-y_0(H_b)$. The constant of proportionality between $\theta$ and vortex displacement is determined by finding the value of $\chi_0$ from the vortex displacement in a static applied field, which is measured by scanning the probe across the vortex \cite{Badea2015}. Though each data point is collected over a time $\sim 1$~s, the dwell time of the vortex at $H_a$ or $H_b$ is only 33~$\mu$s.  Due to the exponential nature of thermal activation, this reduction in dwell time by more than 4 orders of magnitude reduces the thermally surmountable barrier by about $10k_B T$. We roughly estimate this barrier to be on the order of 250~meV in our experiment.  The root mean square (RMS) noise $\sigma \approx 2$~nm sets the displacement resolution. In order to see a pinning site, the jump out of the pinning site must be sufficiently above this noise floor. For a pinning site with half-width $\gamma$ and depth $\delta$, the jump out the pinning site $\delta y \approx \delta / k \gamma$. Setting $\delta y \approx 10$~nm as the minimum clearly visible step, the minimum detectable pinning site has $\delta \approx 1$~eV, for $\gamma=10$~nm.

\begin{figure}[htb]
\centering{}\includegraphics{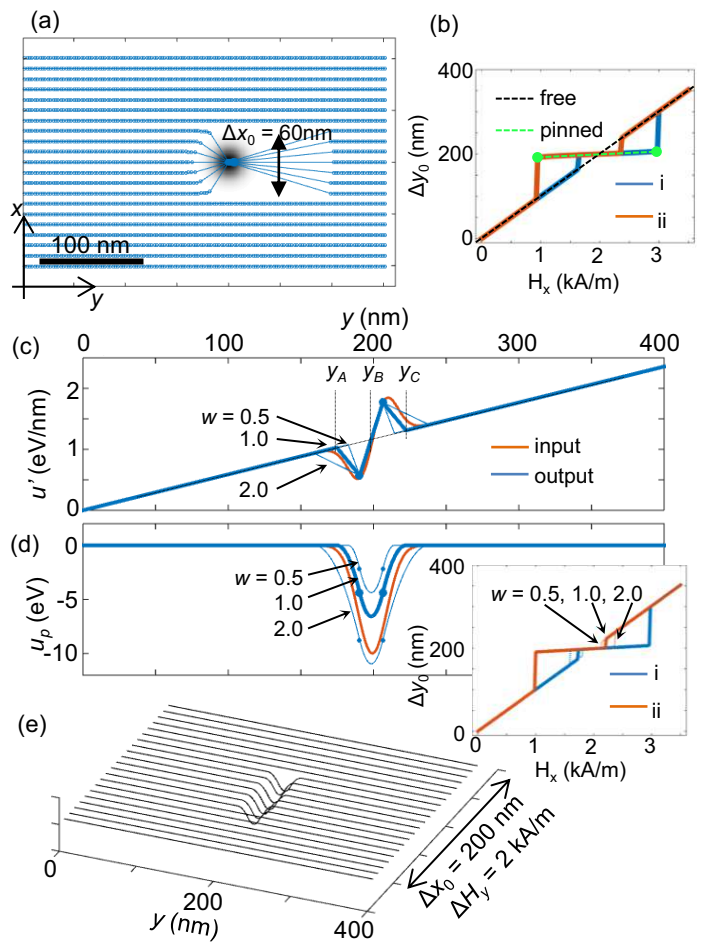}\protect\caption{{\footnotesize{}Demonstration of algorithm. (a) Example Gaussian pinning potential (grayscale.) Simulated vortex paths at fixed $H_y$, and increasing $H_x$ (blue). (b) Simulated vortex displacement on the path through then center of the pinning site for increasing $H_x$ (path i), and decreasing $H_x$ (path ii). (c) Constructed derivative of the effective 1D potential $u'$ at $H_x=0$, at three values of the width parameter $w$ (blue). Derivative of the target potential (red). (d) Target and constructed pinning potentials found by integrating curves in (c). Inset shows hysteresis loops generated from constructed $u_p$. (e) 2D pinning map produced by performing analysis on all sweeps shown in (a). }} 
\label{simfig}
\end{figure}

\begin{figure}[tbp]
\centering{}\includegraphics{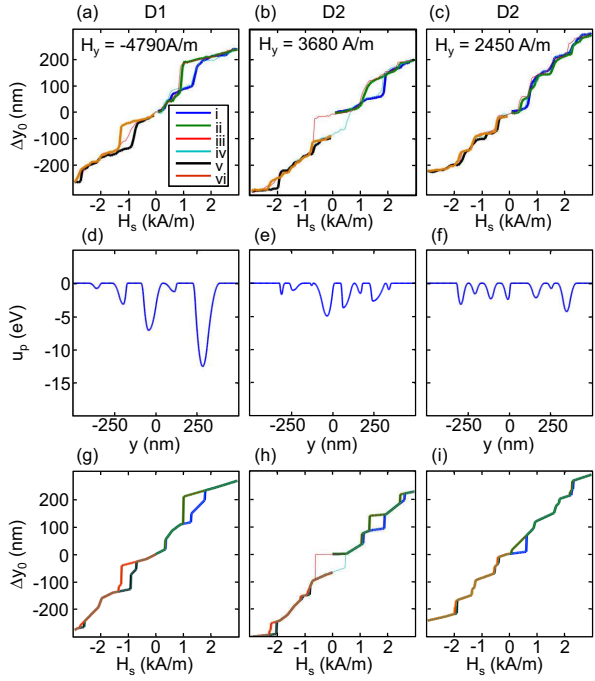}\protect\caption{{\footnotesize{}Extracting 1D pinning potentials from hysteresis loops. (a-c) Vortex displacement $\Delta y_0$ vs. $H_s$ on samples D1 and D2 at different static $H_y$, following the sweep path shown in Fig.~\ref{cycling}a. (d)-(f) Effective 1D pinning potentials $u_p(y)$ extracted from the data shown in (a-c).  (g-i) Simulated vortex displacement $\Delta y_0$ vs. $H_s$ using the potentials $u_p$ in (d-f).}} 
\label{lines}
\end{figure}

We measure the hysteretic vortex displacement $\Delta y_0$ vs. $H_x$ by sweeping the amplitude and offset of the applied square-wave magnetic field as illustrated in Fig.~\ref{cycling}(c), with the blue and red lines indicating the upper level ($H_a$) and lower level ($H_b$) respectively. In each segment (i)-(vi), there is one swept end $H_s$ of the alternating magnetic field, either $H_s=H_a$ or $H_s=H_b$. The resulting measured signal $\theta \propto \Delta y_0$ is shown in Fig.~\ref{cycling}(d) as the difference between the red and blue lines. The red and blue lines show the net vortex displacement from the initial position at $H_a$ and $H_b$. In segment (i), $\theta$ represents the vortex displacement starting from the initial position with field \textit{increasing} to $H_a$. In segment (ii), $\theta$ represents the vortex displacement starting from the endpoint of segment (i), with field \textit{decreasing} to $H_b$.  Thus segments (i) and (ii) yield the initial increase from $H=0$ and decrease back to $H=0$ of a traditional hysteresis loop. Segments (iii) and (iv) reveal the hysteresis loop decreasing and increasing respectively between $H_{max}$ and $H_{min}$.  Finally, the segments (v) and (vi) close the loop, showing the increase to and decrease from $H=0$ on the negative side.  This cycle of $H_a$ and $H_b$ has the advantages of beginning and ending at the same values, and of including some redundancy as a check of repeatability. By plotting vortex displacement $\Delta y_0$ relative to initial position vs. $H_s$, hysteresis loops are formed, as shown in Fig.~\ref{simfig} (simulation) and Fig.~\ref{lines} (data).

To illustrate the algorithm for extracting the pinning potential, we first analyze simulated hysteresis data from a single Gaussian pinning site $U_p(\mathbf{r})$. The pinning site, with half-width $\gamma = 10$~nm, and depth $\delta = 10$~eV, is shown in grayscale in Fig.~\ref{simfig}(a). The vortex position $\mathbf{r}_0$ in the 2D potential $U(\mathbf{r},\mathbf{H})=U_0(\mathbf{r},\mathbf{H})+U_p(\mathbf{r})$ is found using gradient descent. The blue points in Fig.~\ref{simfig}(a) show $\mathbf{r}_0$ at different fixed values of $H_y$ as $H_x$ is increased, moving the vortex from left to right. $H_y$ sets the equilibrium position $x_0$ of the unpinned vortex in the x-direction. At a value of $H_y$ such that the vortex is far from the pinning site, the free vortex moves linearly with $H_x$. In a range $\Delta x_0 = 60$~nm, the vortex approaches sufficiently close to the pinning site for it to become pinned. This length scale is set by the region, extending into the tails of the Gaussian pinning site, over which the gradient of $U_p$ can overcome the gradient of $U_0$. Along paths where pinning occurs, there is a point to the left of the pinning site where the vortex jumps from the free state to near the minimum of $U_p$. As $H_x$ is increased further to $H_x=H_x^+$, the vortex translates within the pinning site to $y=y^+$, and then the vortex jumps out of the pinning site. A sweep of $H_x$ in the opposite direction would look the same, flipped left to right, and would reveal the point $(y^-,H_x^-)$.  

We now turn to the procedure for extracting the effective 1D potentials along the vortex paths shown in Fig.~\ref{simfig}(a). Plotting the simulated $\Delta y_0$ vs. $H_x$ for increasing sweeps (path i, blue) and decreasing sweeps (path ii, red) results in a hysteresis loop, as shown in Fig.~\ref{simfig}(b) for the scan through the center of the pinning site in Fig.~\ref{simfig}(a). We first identify the path of the free vortex, which is drawn in black. This line is set to overlap with any regions of free vortex motion, and to pass through the pinned plateaus. Next, the plateaus corresponding to pinning are identified. In this case, there is only one, and it is shown in green. This procedure is carried out manually. The endpoints of the green line indicate $(y^\pm,H_x^\pm)$. 

The construction of the 1D potential $u(y,H_x)$ is more clearly understood by first constructing its derivative $u'(y,H_x)$. For comparison, the derivative of the potential from the simulation that we are trying to reproduce is shown in Fig.~\ref{simfig}(c) in red. The parabolic free potential has a linear derivative $u'_0 = ky-k\chi_0 H_x$ which is shifted vertically by $H_x$. In the absence of pinning, the equilibrium position $y_0$ occurs where this line crosses zero, at $y_0 = \chi_0 H_x$. The overall slope seen in Fig.~\ref{simfig}(c) illustrates $u'_0(y,0)$. To add the pinning sites, we make use of Eq.~\ref{depinning}. From the requirement that $u'(y^\pm,H_x^\pm)=0$, we obtain the values $u'_p(y^\pm) = -u'_0(y^\pm,H_x^\pm)=-ky^\pm + k\chi_0 H_x^\pm$. These values are added to $u'_0(y^\pm,H_x)$, and are shown at $H_x=0$ as blue dots in Fig.~\ref{simfig}(c). These points are close to the extrema of the target curve in red. All that remains is to connect these points with the free vortex line. We do this by simply connecting $u'(y^\pm,H_x^\pm)$ to the points along the free line at positions $y_A$, $y_B$, and $y_C$ with straight lines, as shown. This corresponds to building the potential $u$ as a quadratic spline. This method is particularly useful because it automatically satisfies the second part of Eq.~\ref{depinning} which requires that $u'(y^\pm,H_x^\pm)$ are local extrema.  We begin by setting $y_B$ so that the points $u'(y^\pm,H_x^\pm)$ are connected linearly. We then set $y_A = y^- - w(y^+ - y^-)$ and $y_C = y^+ - w(y^+ - y^-)$, where the parameter $w$ sets the width of the part of the potential where the vortex position is unstable ($u'' < 0$). Because the vortex position is not stable in the intervals $(y_A,y^-)$ and $(y^+, y_C)$, we don't have direct information about the shape of $u_p$ here. Fig.~\ref{simfig}(c) shows $u'(y,0)$ for three values of $w$. In principle, $w$ could be adjusted for each pinning site if the vortex jumps into it from the free state (as it does in this simulation), by using the position of that jump. To reduce the number of adjustable parameters in this model, however, we find that $w=1$ with the algorithmic adjustment described below acceptably reproduces the observed behavior. 

In setting $y_A$, $y_B$, and $y_C$, we impose two additional constraints: 1. the pinning sites do not overlap, and 2. $u_p(-\infty)=u_p(\infty)$ for each pinning site. We now adjust $y_A$, $y_B$, and $y_C$ chosen above to satisfy constraints 1 and 2. When there is more than one pinning site (which is typical), the pinning potentials are constructed from left to right. It may occur that the desired $y_A <\tilde{y}_C$ where $\tilde{y}_C$ is the value of $y_C$ from the adjacent pinning site to the left. If this occurs, we set $y_A=\tilde{y}_C$, satisfying constraint 1. Furthermore, $y_C$ must not overlap with any (not yet constructed) adjacent pinning site to the right. If $y_C > (y^+ + \tilde{y}^-)/2$, where $\tilde{y}^-$ is the value from the adjacent pinning site to the right, we set  $y_C = (y^+ + \tilde{y}^-)/2$, to ensure that there is room to construct the adjacent pinning potential. Constraint 2 is satisfied when $I=\int u'_p dy = 0$. We first reduce $|I|$ by reducing $w$ for either $y_A$ or $y_C$, down to a minimum of $w=0.2$. (Further reducing $w$ results in unrealistically sharp kinks in the potential.) If $w$ has been reduced all the way to 0.2 on one side, and still $|I|>0$, then $y_B$ is adjusted to set $I=0$. An additional complication can arise when one of the points of the pair $(y^\pm,H_x^\pm)$ is not visible in the data, either because adjacent pinning sites cause this point to be jumped over, or because the point lies outside the scan range. In this case, the unknown point is assumed to lie as symmetrically as possible about the free vortex line from the known point of the pair, consistent with the unknown point not being visible in the data. These methods of adjustment are not unique, but are fairly straightforward, and are found to reliably reproduce the observed pinning behavior. 

Integration yields the effective 1D potential $u_p(y)=\int_0^y u'_p(y') dy'$, as shown for three values of $w$ in Fig.~\ref{simfig}(d). Again, the potential from the simulation is shown in red. To verify that the constructed 1D potential accurately reproduces the simulated vortex displacement, we complete the circle by using a gradient descent method on the reconstructed 1D effective potential to numerically simulate the vortex displacement vs. $H_x$.  The resulting hysteresis loops are shown in the inset to Fig.~\ref{simfig}(d), for the same three values of $w$. Note that while changing $w$ affects the total depth of the pinning potential significantly, the resulting vortex hysteresis is only affected slightly. This is because the pinning behavior is almost entirely determined by the region between $y^-$ and $y^+$ where the vortex position can be stable. These types of hysteresis measurements yield little information about the region where the position is unstable, and therefore do not strongly constrain the total pinning site depth. Instead the measurements do strongly constrain the position, width, and stiffness (maximum gradient) of the pinning sites.

An image reflecting the 2D pinning potential can be generated by constructing the 1D effective potential $u_p$ at each value of $H_y$ as shown in Fig.~\ref{simfig}(e). Plotted in this way, the series of scans roughly shows the shape of the potential in the $x$-direction as well as the $y$-direction. More accurately, the evolution along the series shows how the path of the vortex changes to pass through the pinning site as the minimum of $U_0$ is shifted in the $x$-direction. Along the y-direction, the reconstructed potentials reflect the position, width, and stiffness of the pinning site, while the change in the x-direction shows the range $\Delta x_0$ of the free vortex position over which the vortex can be captured by the pinning site.

Figure~\ref{lines}(a)-(c) shows three characteristic measured $\Delta y_0$ vs. $H_s$ loops. The data shown in (a) was collected from disk (D1) and the data shown in (b) and (c) was collected from disk (D2). All three show multiple jumps in both $\Delta y_0$ and the slope $\chi$ as $H_s$ is swept, corresponding to changes in the pinning state. When the vortex is not pinned, then $\chi=\chi_0$ which is the maximum value of $\chi$ (except for at sudden jumps of $\Delta y_0$). Such a free region can be seen on the left side of Fig.~\ref{lines}(b) bound by $H_{s} = (-1.5,-0.8)$~kA/m. When the vortex is pinned, $\chi$ is reduced, as is the case at $H_s=0$ in both Fig.~\ref{lines}(b) and (c).  The value of $\chi$ indicates the curvature of the pinning potential with a larger $\chi$ corresponding to less curvature. 

From the type of data shown in Fig.~\ref{lines}(a)-(c), we can construct an effective 1D potential that gives rise to the observed loops. First, a small linear background is subtracted from all the data, which arises from a contribution to the signal from the vortex away from the core. That is, even when the core is completely pinned, the magnetization in the surrounding region can be deformed \cite{Burgess2014}.  From the measurements and simulations in Ref. \cite{Badea2015}, we estimate the linear background to be $\approx~0.1\chi_0 H_x$. As described above, jumps in $y_0$ or $\chi$ occur at a set of points $(y_i^\pm,H_{x,i}^\pm)$ that satisfy Eq. 1. These points can be directly read off from the $y_0$ vs. $H_x$ data, and through Eq. 1 specify both the value of $u'_{p,i}$ and that $u'_{p,i}$ is a local maximum or minimum at that value of $y$. The effective potential $u(y,H_x)$ is then constructed as demonstrated above with simulated data. 

The effective 1D potentials $u_p=\sum{u_{p,i}}$ constructed from the data in Fig.~\ref{lines}(a)-(c) are shown in Fig.~\ref{lines}(d)-(f).  Again, note that the quantities specified by the measured data are values and positions of maximum or minimum $u'_{p,i}$, and hence the width and stiffness of the pinning sites. Since, for simplicity, we have used a quadratic spline to connect these known points, the particular shape of the pinning potential, including the depth of the wells, should be understood to be qualitative.  Nevertheless, the constructed potential does contain the necessary detail to reproduce the observed data.  To verify this, as with the simulated data, we use a gradient descent method on the reconstructed 1D effective potential to numerically simulate the vortex displacement vs. $H_x$.  The resulting simulated hysteresis loops are shown in Fig.~\ref{lines}(g)-(i).  In each case, the salient features of the data are reproduced. The simulation does not take thermal effects into account, so does not capture some smearing of the jumps or reduction of hysteresis for small pinning sites.

\begin{figure}[tbp]
\centering{}\includegraphics{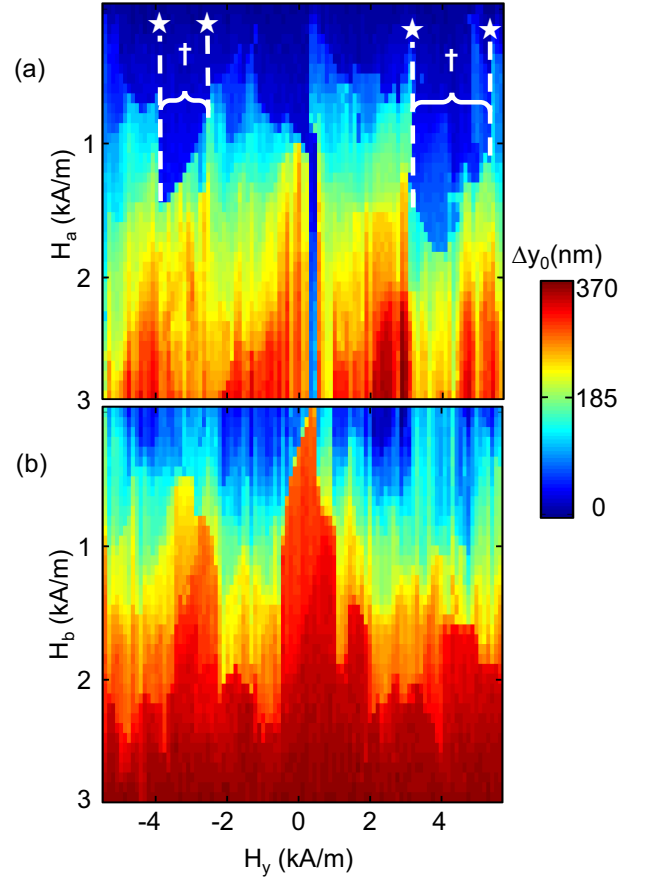}\protect\caption{{\footnotesize{}2D hysteresis maps.(a) and (b) Path (i) and (ii) of vortex displacement $\Delta y_0$ vs. $H_a$ and $H_b$, as static field $H_y$ shifts the free vortex position in the $x$-direction. Examples of regions of $H_y$ where the vortex is trapped by the same pinning site are indicated with $\dagger$, separated by transitions labelled $\star$. }} 
\label{mokemap}
\end{figure}

We map the pinning potential in 2D by collecting a series of hysteresis loops of $\Delta y_0$ as above, incrementing the unpinned equilibrium position $\Delta x_0$ in the $x$-direction, using a static field $H_y$. Figure~\ref{mokemap}(a) and (b) show path (i) and (ii) of the hysteresis loops vs. $H_x$ over the range $H_y=-5570$ to $5570$~A/m, which corresponds to a translation $\Delta x_0 \approx 1~\mu$m. As in Fig.~\ref{lines}, jumps in $\Delta y_0$ with increasing $H_x$ yield the critical fields $H^\pm_{x,i}$ as the vortex jumps into or out of the $i$th pinning site.  Differences between Fig.~\ref{mokemap}(a) and (b) reveal hysteresis in $\Delta y_0$. These maps show features (e.g. labelled with $\dagger$) with qualitatively similar behavior over a range of $H_y$, which end in sudden transitions (e.g. labelled with $\star$).  A particular $\dagger$ feature is associated with a particular pinning potential $U_{p,i}$. There is a range of $H_y$ over which the vortex is captured by $U_{p,i}$ as $H_x$ is swept. Over this range, the 2D path of the vortex through $U_{p,i}$ shifts, resulting in a continuous change in the corresponding 1D effective potential $u_{p,i}$. As a result, the critical fields $H^\pm_{x,i}$ change over that range, giving rise to the curvature of the $\dagger$ features.  At some value of $H_y$, the path of the vortex does not approach $U_{p,i}$ sufficiently to capture the vortex, and instead the vortex is captured by a different pinning site, or remains free at the minimum of $U_0$.  At these critical values of $H_y$, a sudden change in the $\Delta y_0$ scans appears, such as at the $\star$ features.

\begin{figure*}[tbp]
\centering{}\includegraphics{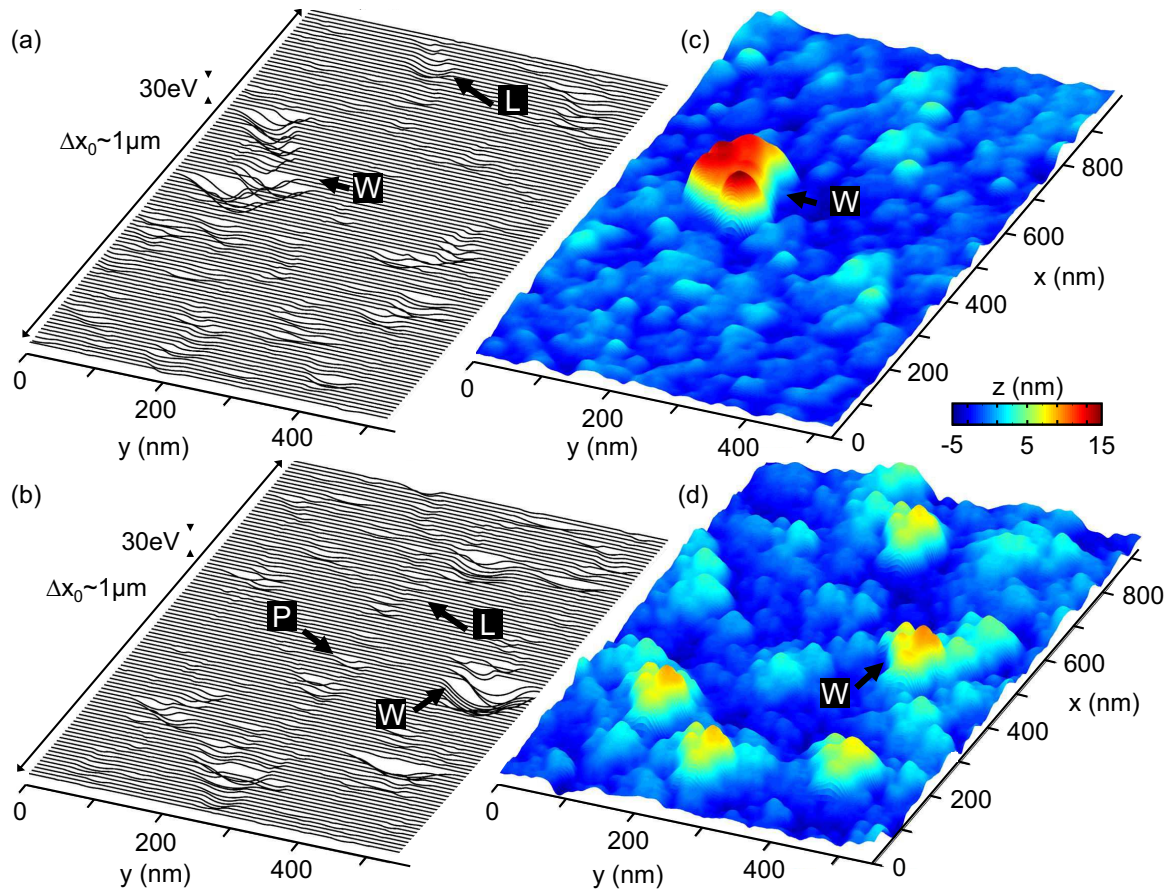}\protect\caption{{\footnotesize{}2D maps of the pinning potential. (a) and (b) Maps of pinning potential from samples D1 and D2 produced from series of 1D effective potentials with static $H_y = -5570$ to $5570$~A/m in 100 steps. Examples of three types of pinning sites are labelled $\mathbf{P}$,$\mathbf{W}$, and $\mathbf{L}$. (c) and (d) Topographical maps of the surfaces of samples D1 and D2 taken by means of tapping mode AFM. The z-axis is exaggerated to make features more easily visible.  }} 
\label{pinningmap}
\end{figure*}

We now generate the 2D pinning map as for the simulated data in Fig.~\ref{simfig}(e), by constructing the 1D effective potential $u_p$ at each value of $H_y$ as shown in Fig.~\ref{pinningmap}(a) and (b) for disks D1 and D2. The series of scans cover $H_y=-5570$ to $5570$~A/m in 100 steps. Each step of $H_y$ shifts the minimum of $U_0$ by $\Delta x_0 \approx 10$~nm. The maps show features that persist across a range $\Delta H_y$, or a corresponding range $\Delta x_0$. These features correspond to pinning sites $U_{p,i}$, with $\Delta x_0$ limited by the disappearance of the local minimum or by the vortex path not approaching sufficiently close to be captured. Topographical images of the surfaces of D1 and D2 are shown in Fig.~\ref{pinningmap}(c) and (d) taken by tapping mode AFM over a $3 \times 3$~$\mu$m region. The images were cropped and centered on the disk to enclose approximately the same region shown in Fig.~\ref{pinningmap}(a) and (b).

We categorize the pinning sites in Fig.~\ref{pinningmap}(a) and (b) into three general types, with examples labelled $\bf{P}$, $\bf{W}$, and $\bf{L}$. The most visible features in the pinning maps are the $\bf{W}$ (``wide'') features which are characterized by half-widths (along the y-direction) $\gamma \sim 75$~nm, depths $\delta \sim 20$~eV, and $\Delta x_0 \sim 200$~nm. The spatial extent of these features in both $x-$ and $y-$directions suggests an extended pinning site, significantly larger than the vortex core. Indeed, the $\bf{W}$ pinning sites are well correlated with the several large (height $\approx 10$~nm) bumps visible in the AFM data. There is one such prominent feature in disk D1, and about five such features in disk D2. Pinning behavior similar to $\bf{W}$ features has been previously observed in artificially created pinning sites \cite{Rahm2003, Rahm2004, Rahm2004b, Uhlig2005}. Although no attempt was made to intentionally create these large pinning sites, the $\bf{W}$ sites are believed to be a product of residual resist left on the Au stripline during the lithographic stage in the fabrication process. 

 $\bf{P}$ (``point'') features are characterized by half-widths $\gamma \sim 10$~nm, depths $\delta \sim 1$~eV, and $\Delta x_0\sim 100$~nm. These are consistent with point-like defects in the Py, where $\gamma$ is set by the vortex core half-width, and $\Delta x_0$ is set by the full range over which the vortex core has significant out-of-plane magnetization \cite{Hollinger2003}, as in the simulated pinning site in Fig.~\ref{simfig}. Similar pinning potentials have been previously inferred using torsional microresonators \cite{Burgess2013}, or by mapping the vortex gyrotropic frequency \cite{Chen2012, Chen2012b, Compton2006, Compton2010}.
 
In contrast to the symmetric nature of both $\bf{W}$ and $\bf{P}$ features, $\bf{L}$ (``line'') features appear to arise from line-like defects. These features have $\gamma \sim 10$~nm, $\delta \sim 10$~eV, and very large $\Delta x_0 \sim 300$~nm or more.  Here, $\gamma$ is consistent with a point-like defect in the $y$-direction, but the large value of $\Delta x_0$ indicates significant spatial extent in the $x$ direction. Note that this measurement is only sensitive to $\bf{L}$ features oriented mainly along the $x$-axis:  for a line defect aligned close to the $y$-axis, the vortex would be free to move along the potential minimum over the entire hysteresis scan, and would look essentially like the unpinned case.  Line defects at intermediate angles would likely give rise to jumps in the $\Delta y_0$ scan as the vortex goes into or out of the pinning potential separated by free motion as it translates along the line. Such behavior may not be distinguishable from that of a pair of $\bf{P}$ features. Grain boundaries are theorized to be a possible source of pinning which might produce $\bf{L}$ type features.\cite{Uhlig2005} To determine the grain size we used a grain counting algorithm provided by a commercial AFM analysis program (PicoView), on regions of the sample away from any $\bf{W}$ features. For samples D1 and D2 the average grain size was determined to be $\sigma_{g} = 46 \pm 2$~nm. Because both the length of the $\bf{L}$ features and the spacing between them is much greater than the typical grain size, we can at least conclude that not every grain boundary gives rise to an $\bf{L}$ feature in the image. It is possible that the pinning created by a typical grain boundary is too weak to observe here, and only grain boundaries that result in particularly long, deep trenches in the topography give rise to $\bf{L}$ features. Such trenches can perhaps be seen in the AFM data, but the correlation between the topography and the pinning map is not sufficient to make any positive identification.

In this work, we have developed a technique for mapping the effective potential that pins a ferromagnetic vortex as it moves through a film. The method is based on differential magneto-optical microscopy, and is capable of resolving features with depth $\sim 1$~eV and width $\sim 10$~nm. By using a magnetic field modulated at $15$~kHz, the effects of thermal activation out of pinning sites are suppressed. From the hysteretic displacement of the vortex vs. magnetic field, we obtain an effective 1D potential along a path approximately in the y-direction. Shifting the position of this line in the x-direction and repeating this procedure maps out how the pinning potential changes across the film. As a next step, fast AC magnetic fields could be applied in an arbitrary in-plane direction for full 2D mapping of the pinning potential. We have shown pinning maps in two permalloy disks, and categorize the observed pinning features into three categories -- large, roughly symmetric features, point-like features, and line-like features. By comparison to AFM images of these films in same region as the pinning maps, the largest pinning features are found to be well-correlated with $\sim 10$-nm-high features in the topography, serving as a check on the reliability of the pinning maps. The origin of the point- and line-like features observed in the pinning maps is still an open question, though the AFM maps suggest there may be topographic features associated with these pinning sites as well. Further characterization of the defects present in the film will allow more detailed comparison to the pinning maps.

The technique presented here provides a route towards making connections between the microscopic defects in a FM film, and how those defects cause domain wall pinning. This technique may be used in future work to understand how the fabrication process or post-fabrication treatment can be used to control the degree and type of pinning present in a sample.

\begin{acknowledgements}
This work was supported by DOE, award No. DE-SC008148. 
\end{acknowledgements}

\begin{thebibliography}{24}%
\makeatletter
\providecommand \@ifxundefined [1]{%
 \@ifx{#1\undefined}
}%
\providecommand \@ifnum [1]{%
 \ifnum #1\expandafter \@firstoftwo
 \else \expandafter \@secondoftwo
 \fi
}%
\providecommand \@ifx [1]{%
 \ifx #1\expandafter \@firstoftwo
 \else \expandafter \@secondoftwo
 \fi
}%
\providecommand \natexlab [1]{#1}%
\providecommand \enquote  [1]{``#1''}%
\providecommand \bibnamefont  [1]{#1}%
\providecommand \bibfnamefont [1]{#1}%
\providecommand \citenamefont [1]{#1}%
\providecommand \href@noop [0]{\@secondoftwo}%
\providecommand \href [0]{\begingroup \@sanitize@url \@href}%
\providecommand \@href[1]{\@@startlink{#1}\@@href}%
\providecommand \@@href[1]{\endgroup#1\@@endlink}%
\providecommand \@sanitize@url [0]{\catcode `\\12\catcode `\$12\catcode
  `\&12\catcode `\#12\catcode `\^12\catcode `\_12\catcode `\%12\relax}%
\providecommand \@@startlink[1]{}%
\providecommand \@@endlink[0]{}%
\providecommand \url  [0]{\begingroup\@sanitize@url \@url }%
\providecommand \@url [1]{\endgroup\@href {#1}{\urlprefix }}%
\providecommand \urlprefix  [0]{URL }%
\providecommand \Eprint [0]{\href }%
\providecommand \doibase [0]{http://dx.doi.org/}%
\providecommand \selectlanguage [0]{\@gobble}%
\providecommand \bibinfo  [0]{\@secondoftwo}%
\providecommand \bibfield  [0]{\@secondoftwo}%
\providecommand \translation [1]{[#1]}%
\providecommand \BibitemOpen [0]{}%
\providecommand \bibitemStop [0]{}%
\providecommand \bibitemNoStop [0]{.\EOS\space}%
\providecommand \EOS [0]{\spacefactor3000\relax}%
\providecommand \BibitemShut  [1]{\csname bibitem#1\endcsname}%
\let\auto@bib@innerbib\@empty
\bibitem [{\citenamefont {Barkhausen}(1916)}]{Barkhausen1916}%
  \BibitemOpen
  \bibfield  {author} {\bibinfo {author} {\bibfnamefont {H.}~\bibnamefont
  {Barkhausen}},\ }\bibfield  {title} {\enquote {\bibinfo {title} {Zwei mit
  hilfe der neuen verst{\"a}rker entdeckte erscheinungen},}\ }\href@noop {}
  {\bibfield  {journal} {\bibinfo  {journal} {Physik Zeitschrift}\ }\textbf
  {\bibinfo {volume} {20}},\ \bibinfo {pages} {401--403} (\bibinfo {year}
  {1916})}\BibitemShut {NoStop}%
\bibitem [{\citenamefont {Hubert}\ and\ \citenamefont
  {Sch{\"a}fer}(1996)}]{Hubert1998}%
  \BibitemOpen
  \bibfield  {author} {\bibinfo {author} {\bibfnamefont {Alex}\ \bibnamefont
  {Hubert}}\ and\ \bibinfo {author} {\bibfnamefont {Rudolf}\ \bibnamefont
  {Sch{\"a}fer}},\ }\href@noop {} {\emph {\bibinfo {title} {Magnetic Domains:
  The Analysis of Magnetic Microstructures}}}\ (\bibinfo  {publisher}
  {Springer-Verlag Berlin Heidelberg},\ \bibinfo {address} {Berlin},\ \bibinfo
  {year} {1996})\BibitemShut {NoStop}%
\bibitem [{\citenamefont {Durin}\ and\ \citenamefont
  {Zapperi}(2006)}]{Durin2006}%
  \BibitemOpen
  \bibfield  {author} {\bibinfo {author} {\bibfnamefont {Gianfranco}\
  \bibnamefont {Durin}}\ and\ \bibinfo {author} {\bibfnamefont {Stefano}\
  \bibnamefont {Zapperi}},\ }\bibfield  {title} {\enquote {\bibinfo {title}
  {Chapter 3 - the barkhausen effect},}\ }in\ \href {\doibase
  http://dx.doi.org/10.1016/B978-012480874-4/50014-2} {\emph {\bibinfo
  {booktitle} {The Science of Hysteresis}}},\ \bibinfo {editor} {edited by\
  \bibinfo {editor} {\bibfnamefont {Giorgio BertottiIsaak~D.}\ \bibnamefont
  {Mayergoyz}}}\ (\bibinfo  {publisher} {Academic Press},\ \bibinfo {address}
  {Oxford},\ \bibinfo {year} {2006})\ pp.\ \bibinfo {pages} {181 --
  267}\BibitemShut {NoStop}%
\bibitem [{\citenamefont {Cowburn}\ \emph {et~al.}(1999)\citenamefont
  {Cowburn}, \citenamefont {Koltsov}, \citenamefont {Adeyeye},\ and\
  \citenamefont {Welland}}]{Cowburn1999}%
  \BibitemOpen
  \bibfield  {author} {\bibinfo {author} {\bibfnamefont {R.P.}\ \bibnamefont
  {Cowburn}}, \bibinfo {author} {\bibfnamefont {D.K.}\ \bibnamefont {Koltsov}},
  \bibinfo {author} {\bibfnamefont {A.O.}\ \bibnamefont {Adeyeye}}, \bibinfo {author} {\bibfnamefont {M.E.}\ \bibnamefont {Welland}}\ and\
  \bibinfo {author} {\bibfnamefont {D.M.}\ \bibnamefont {Tricker}},\ }\bibfield
   {title} {\enquote {\bibinfo {title} {Single-domain circular nanomagnets},}\
  }\href {\doibase 10.1103/PhysRevLett.83.1042} {\bibfield  {journal} {\bibinfo
   {journal} {Physical Review Letters}\ }\textbf {\bibinfo {volume} {83}},\
  \bibinfo {pages} {1042--1045} (\bibinfo {year} {1999})}\BibitemShut {NoStop}%
\bibitem [{\citenamefont {Shinjo}\ \emph {et~al.}(2000)\citenamefont {Shinjo},
  \citenamefont {Okuno}, \citenamefont {Hassdorf}, \citenamefont {Shigeto},\
  and\ \citenamefont {Ono}}]{Shinjo2000}%
  \BibitemOpen
  \bibfield  {author} {\bibinfo {author} {\bibfnamefont {T.}~\bibnamefont
  {Shinjo}}, \bibinfo {author} {\bibfnamefont {T.}~\bibnamefont {Okuno}},
  \bibinfo {author} {\bibfnamefont {R.}~\bibnamefont {Hassdorf}}, \bibinfo
  {author} {\bibfnamefont {K.}~\bibnamefont {Shigeto}}, \ and\ \bibinfo
  {author} {\bibfnamefont {T.}~\bibnamefont {Ono}},\ }\bibfield  {title}
  {\enquote {\bibinfo {title} {Magnetic vortex core observation in circular
  dots of permalloy},}\ }\href {\doibase 10.1126/science.289.5481.930}
  {\bibfield  {journal} {\bibinfo  {journal} {Science}\ }\textbf {\bibinfo
  {volume} {289}},\ \bibinfo {pages} {930--932} (\bibinfo {year}
  {2000})}\BibitemShut {NoStop}%
\bibitem [{\citenamefont {Wachowiak}\ \emph {et~al.}(2002)\citenamefont
  {Wachowiak}, \citenamefont {Wiebe}, \citenamefont {Bode}, \citenamefont
  {Pietzsch}, \citenamefont {Morgenstern},\ and\ \citenamefont
  {Wiesendanger}}]{Wachowiak2002}%
  \BibitemOpen
  \bibfield  {author} {\bibinfo {author} {\bibfnamefont {A.}~\bibnamefont
  {Wachowiak}}, \bibinfo {author} {\bibfnamefont {J.}~\bibnamefont {Wiebe}},
  \bibinfo {author} {\bibfnamefont {M.}~\bibnamefont {Bode}}, \bibinfo {author}
  {\bibfnamefont {O.}~\bibnamefont {Pietzsch}}, \bibinfo {author}
  {\bibfnamefont {M.}~\bibnamefont {Morgenstern}}, \ and\ \bibinfo {author}
  {\bibfnamefont {R.}~\bibnamefont {Wiesendanger}},\ }\bibfield  {title}
  {\enquote {\bibinfo {title} {Direct observation of internal spin structure of
  magnetic vortex cores},}\ }\href {\doibase 10.1126/science.1075302}
  {\bibfield  {journal} {\bibinfo  {journal} {Science}\ }\textbf {\bibinfo
  {volume} {298}},\ \bibinfo {pages} {577--580} (\bibinfo {year}
  {2002})}\BibitemShut {NoStop}%
\bibitem [{\citenamefont {H{\"o}llinger}\ \emph {et~al.}(2003)\citenamefont
  {H{\"o}llinger}, \citenamefont {Killinger},\ and\ \citenamefont
  {Krey}}]{Hollinger2003}%
  \BibitemOpen
  \bibfield  {author} {\bibinfo {author} {\bibfnamefont {R.}~\bibnamefont
  {H{\"o}llinger}}, \bibinfo {author} {\bibfnamefont {A.}~\bibnamefont
  {Killinger}}, \ and\ \bibinfo {author} {\bibfnamefont {U.}~\bibnamefont
  {Krey}},\ }\bibfield  {title} {\enquote {\bibinfo {title} {Statics and fast
  dynamics of nanomagnets with vortex structure},}\ }\href {\doibase
  10.1016/S0304-8853(02)01471-3} {\bibfield  {journal} {\bibinfo  {journal}
  {Journal of Magnetism and Mangetic Materials}\ }\textbf {\bibinfo {volume}
  {261}},\ \bibinfo {pages} {178--189} (\bibinfo {year} {2003})}\BibitemShut
  {NoStop}%
\bibitem [{\citenamefont {Burgess}\ \emph {et~al.}(2013)\citenamefont
  {Burgess}, \citenamefont {Fraser}, \citenamefont {{Fani Sani}}, \citenamefont
  {Vick}, \citenamefont {Hauer}, \citenamefont {Davis},\ and\ \citenamefont
  {Freeman}}]{Burgess2013}%
  \BibitemOpen
  \bibfield  {author} {\bibinfo {author} {\bibfnamefont {J.A.J.}\ \bibnamefont
  {Burgess}}, \bibinfo {author} {\bibfnamefont {A.E.}\ \bibnamefont {Fraser}},
  \bibinfo {author} {\bibfnamefont {F.}~\bibnamefont {{Fani Sani}}}, \bibinfo
  {author} {\bibfnamefont {D.}~\bibnamefont {Vick}}, \bibinfo {author}
  {\bibfnamefont {B.D.}\ \bibnamefont {Hauer}}, \bibinfo {author}
  {\bibfnamefont {J.P.}\ \bibnamefont {Davis}}, \ and\ \bibinfo {author}
  {\bibfnamefont {M.R.}\ \bibnamefont {Freeman}},\ }\bibfield  {title}
  {\enquote {\bibinfo {title} {Quantitative magneto-mechanical detection and
  control of the barkhausen effect},}\ }\href {\doibase
  10.1126/science.1231390} {\bibfield  {journal} {\bibinfo  {journal}
  {Science}\ }\textbf {\bibinfo {volume} {339}},\ \bibinfo {pages} {1051--4}
  (\bibinfo {year} {2013})}\BibitemShut {NoStop}%
\bibitem [{\citenamefont {Min}\ \emph {et~al.}(2010)\citenamefont {Min},
  \citenamefont {McMichael}, \citenamefont {Donahue}, \citenamefont {Miltat},\
  and\ \citenamefont {Stiles}}]{Min2010}%
  \BibitemOpen
  \bibfield  {author} {\bibinfo {author} {\bibfnamefont {H.}~\bibnamefont
  {Min}}, \bibinfo {author} {\bibfnamefont {R.D.}\ \bibnamefont {McMichael}},
  \bibinfo {author} {\bibfnamefont {M.J.}\ \bibnamefont {Donahue}}, \bibinfo
  {author} {\bibfnamefont {J.}~\bibnamefont {Miltat}}, \ and\ \bibinfo {author}
  {\bibfnamefont {M.D.}\ \bibnamefont {Stiles}},\ }\bibfield  {title} {\enquote
  {\bibinfo {title} {Effects of disorder and internal dynamics on vortex wall
  propagation},}\ }\href {\doibase 10.1103/PhysRevLett.104.217201} {\bibfield
  {journal} {\bibinfo  {journal} {Physical Review Letters}\ }\textbf {\bibinfo
  {volume} {104}},\ \bibinfo {pages} {217201} (\bibinfo {year}
  {2010})}\BibitemShut {NoStop}%
\bibitem [{\citenamefont {Parkin}\ \emph {et~al.}(1998)\citenamefont {Parkin},
  \citenamefont {Hayashi},\ and\ \citenamefont {Thomas}}]{Parkin2008}%
  \BibitemOpen
  \bibfield  {author} {\bibinfo {author} {\bibfnamefont {S.}~\bibnamefont
  {Parkin}}, \bibinfo {author} {\bibfnamefont {M.}~\bibnamefont {Hayashi}}, \
  and\ \bibinfo {author} {\bibfnamefont {L.}~\bibnamefont {Thomas}},\
  }\bibfield  {title} {\enquote {\bibinfo {title} {Magnetic domain-wall
  racetrack memory},}\ }\href {\doibase 10.1126/science.1145799} {\bibfield
  {journal} {\bibinfo  {journal} {Science}\ }\textbf {\bibinfo {volume}
  {320}},\ \bibinfo {pages} {190} (\bibinfo {year} {1998})}\BibitemShut
  {NoStop}%
\bibitem [{\citenamefont {Rahm}\ \emph {et~al.}(2003)\citenamefont {Rahm},
  \citenamefont {Biberger}, \citenamefont {Umansky},\ and\ \citenamefont
  {Weiss}}]{Rahm2003}%
  \BibitemOpen
  \bibfield  {author} {\bibinfo {author} {\bibfnamefont {M.}~\bibnamefont
  {Rahm}}, \bibinfo {author} {\bibfnamefont {J.}~\bibnamefont {Biberger}},
  \bibinfo {author} {\bibfnamefont {V.}~\bibnamefont {Umansky}}, \ and\
  \bibinfo {author} {\bibfnamefont {D.}~\bibnamefont {Weiss}},\ }\bibfield
  {title} {\enquote {\bibinfo {title} {Vortex pinning at individual defects in
  magnetic nanodisks},}\ }\href {\doibase 10.1063/1.1558255} {\bibfield
  {journal} {\bibinfo  {journal} {Journal of Applied Sciences}\ }\textbf
  {\bibinfo {volume} {93}},\ \bibinfo {pages} {7429--7431} (\bibinfo {year}
  {2003})}\BibitemShut {NoStop}%
\bibitem [{\citenamefont {Rahm}\ \emph
  {et~al.}(2004{\natexlab{a}})\citenamefont {Rahm}, \citenamefont {Stahl},
  \citenamefont {Wegscheider},\ and\ \citenamefont {Weiss}}]{Rahm2004}%
  \BibitemOpen
  \bibfield  {author} {\bibinfo {author} {\bibfnamefont {M.}~\bibnamefont
  {Rahm}}, \bibinfo {author} {\bibfnamefont {J.}~\bibnamefont {Stahl}},
  \bibinfo {author} {\bibfnamefont {W.}~\bibnamefont {Wegscheider}}, \ and\
  \bibinfo {author} {\bibfnamefont {D.}~\bibnamefont {Weiss}},\ }\bibfield
  {title} {\enquote {\bibinfo {title} {Multistable switching due to magnetic
  vortices pinned at artificial pinning sites},}\ }\href {\doibase
  10.1063/1.1785281} {\bibfield  {journal} {\bibinfo  {journal} {Applied
  Physics Letters}\ }\textbf {\bibinfo {volume} {85}},\ \bibinfo {pages} {1553}
  (\bibinfo {year} {2004}{\natexlab{a}})}\BibitemShut {NoStop}%
\bibitem [{\citenamefont {Rahm}\ \emph
  {et~al.}(2004{\natexlab{b}})\citenamefont {Rahm}, \citenamefont
  {H{\"o}llinger}, \citenamefont {Umansky},\ and\ \citenamefont
  {D.Weiss}}]{Rahm2004b}%
  \BibitemOpen
  \bibfield  {author} {\bibinfo {author} {\bibfnamefont {M.}~\bibnamefont
  {Rahm}}, \bibinfo {author} {\bibfnamefont {R.}~\bibnamefont {H{\"o}llinger}},
  \bibinfo {author} {\bibfnamefont {V.}~\bibnamefont {Umansky}}, \ and\
  \bibinfo {author} {\bibnamefont {D.Weiss}},\ }\bibfield  {title} {\enquote
  {\bibinfo {title} {Influence of point defects on magnetic vortex
  structures},}\ }\href {\doibase 10.1063/1.1667448} {\bibfield  {journal}
  {\bibinfo  {journal} {Journal of Applied Physics}\ }\textbf {\bibinfo
  {volume} {95}},\ \bibinfo {pages} {6708} (\bibinfo {year}
  {2004}{\natexlab{b}})}\BibitemShut {NoStop}%
\bibitem [{\citenamefont {Uhlig}\ \emph {et~al.}(2005)\citenamefont {Uhlig},
  \citenamefont {Rahm}, \citenamefont {Dietrich}, \citenamefont {Hollinger},
  \citenamefont {Heumann}, \citenamefont {Weiss},\ and\ \citenamefont
  {Zweck}}]{Uhlig2005}%
  \BibitemOpen
  \bibfield  {author} {\bibinfo {author} {\bibfnamefont {T.}~\bibnamefont
  {Uhlig}}, \bibinfo {author} {\bibfnamefont {M.}~\bibnamefont {Rahm}},
  \bibinfo {author} {\bibfnamefont {C.}~\bibnamefont {Dietrich}}, \bibinfo
  {author} {\bibfnamefont {R.}~\bibnamefont {Hollinger}}, \bibinfo {author}
  {\bibfnamefont {M.}~\bibnamefont {Heumann}}, \bibinfo {author} {\bibfnamefont
  {D.}~\bibnamefont {Weiss}}, \ and\ \bibinfo {author} {\bibfnamefont
  {J.}~\bibnamefont {Zweck}},\ }\bibfield  {title} {\enquote {\bibinfo {title}
  {Shifting and pinning of a magnetic vortex core in a parmalloy dot by a
  magnetic field},}\ }\href {\doibase 10.1103/PhysRevLett.95.237205} {\bibfield
   {journal} {\bibinfo  {journal} {Physical Review Letters}\ }\textbf {\bibinfo
  {volume} {95}},\ \bibinfo {pages} {237205} (\bibinfo {year}
  {2005})}\BibitemShut {NoStop}%
\bibitem [{\citenamefont {Kuepper}\ \emph {et~al.}(2007)\citenamefont
  {Kuepper}, \citenamefont {Bishoff}, \citenamefont {Akhmadaliev},
  \citenamefont {Fassbender}, \citenamefont {Stoll}, \citenamefont {Chou},
  \citenamefont {Puzic}, \citenamefont {Fauth}, \citenamefont {Dolgos},
  \citenamefont {Sch{\"u}ltz}, \citenamefont {Waeyenberge}, \citenamefont
  {Tyliszczak}, \citenamefont {Neudecker}, \citenamefont {Woltersdorf},\ and\
  \citenamefont {Back}}]{Kuepper2007}%
  \BibitemOpen
  \bibfield  {author} {\bibinfo {author} {\bibfnamefont {K.}~\bibnamefont
  {Kuepper}}, \bibinfo {author} {\bibfnamefont {L.}~\bibnamefont {Bishoff}},
  \bibinfo {author} {\bibfnamefont {Ch.}\ \bibnamefont {Akhmadaliev}}, \bibinfo
  {author} {\bibfnamefont {J.}~\bibnamefont {Fassbender}}, \bibinfo {author}
  {\bibfnamefont {H.}~\bibnamefont {Stoll}}, \bibinfo {author} {\bibfnamefont
  {K.}~\bibnamefont {Chou}}, \bibinfo {author} {\bibfnamefont {A.}~\bibnamefont
  {Puzic}}, \bibinfo {author} {\bibfnamefont {K.}~\bibnamefont {Fauth}},
  \bibinfo {author} {\bibfnamefont {D.}~\bibnamefont {Dolgos}}, \bibinfo
  {author} {\bibfnamefont {G.}~\bibnamefont {Sch{\"u}ltz}}, \bibinfo {author}
  {\bibfnamefont {V.}~\bibnamefont {Waeyenberge}}, \bibinfo {author}
  {\bibfnamefont {T.}~\bibnamefont {Tyliszczak}}, \bibinfo {author}
  {\bibfnamefont {I.}~\bibnamefont {Neudecker}}, \bibinfo {author}
  {\bibfnamefont {G.}~\bibnamefont {Woltersdorf}}, \ and\ \bibinfo {author}
  {\bibfnamefont {C.}~\bibnamefont {Back}},\ }\bibfield  {title} {\enquote
  {\bibinfo {title} {Vortex dynamics in permalloy disks with artificial
  defects: Suppression of the gyrotropic mode},}\ }\href {\doibase
  10.1063/1.2437710} {\bibfield  {journal} {\bibinfo  {journal} {Applied
  Physics Letters}\ }\textbf {\bibinfo {volume} {90}},\ \bibinfo {pages}
  {062506} (\bibinfo {year} {2007})}\BibitemShut {NoStop}%
\bibitem [{\citenamefont {Chen}\ \emph
  {et~al.}(2012{\natexlab{a}})\citenamefont {Chen}, \citenamefont
  {Galkiewicz},\ and\ \citenamefont {Crowell}}]{Chen2012}%
  \BibitemOpen
  \bibfield  {author} {\bibinfo {author} {\bibfnamefont {T.Y.}\ \bibnamefont
  {Chen}}, \bibinfo {author} {\bibfnamefont {A.T.}\ \bibnamefont {Galkiewicz}},
  \ and\ \bibinfo {author} {\bibfnamefont {P.A.}\ \bibnamefont {Crowell}},\
  }\bibfield  {title} {\enquote {\bibinfo {title} {Phase diagram of magnetic
  vortex dynamics},}\ }\href {\doibase 10.1103/PhysRevB.85.180406} {\bibfield
  {journal} {\bibinfo  {journal} {Physical Review B}\ }\textbf {\bibinfo
  {volume} {85}},\ \bibinfo {pages} {180406} (\bibinfo {year}
  {2012}{\natexlab{a}})}\BibitemShut {NoStop}%
\bibitem [{\citenamefont {Chen}\ \emph
  {et~al.}(2012{\natexlab{b}})\citenamefont {Chen}, \citenamefont {Erickson},
  \citenamefont {Crowell},\ and\ \citenamefont {C.Leighton}}]{Chen2012b}%
  \BibitemOpen
  \bibfield  {author} {\bibinfo {author} {\bibfnamefont {T.Y.}\ \bibnamefont
  {Chen}}, \bibinfo {author} {\bibfnamefont {M.J.}\ \bibnamefont {Erickson}},
  \bibinfo {author} {\bibfnamefont {P.A.}\ \bibnamefont {Crowell}}, \ and\
  \bibinfo {author} {\bibnamefont {C.Leighton}},\ }\bibfield  {title} {\enquote
  {\bibinfo {title} {Surface roughness dominated pinning mechanisms of magnetic
  vortices in soft ferromagnetic films},}\ }\href {\doibase
  10.1103/PhysRevLett.109.097202} {\bibfield  {journal} {\bibinfo  {journal}
  {Physical Review Letters}\ }\textbf {\bibinfo {volume} {109}},\ \bibinfo
  {pages} {097202} (\bibinfo {year} {2012}{\natexlab{b}})}\BibitemShut
  {NoStop}%
\bibitem [{\citenamefont {Compton}\ and\ \citenamefont
  {Crowell}(2006)}]{Compton2006}%
  \BibitemOpen
  \bibfield  {author} {\bibinfo {author} {\bibfnamefont {R.L.}\ \bibnamefont
  {Compton}}\ and\ \bibinfo {author} {\bibfnamefont {P.A.}\ \bibnamefont
  {Crowell}},\ }\bibfield  {title} {\enquote {\bibinfo {title} {Dynamics of a
  pinned magnetic vortex},}\ }\href {\doibase 10.1103/PhysRevLett.97.137202}
  {\bibfield  {journal} {\bibinfo  {journal} {Physical Review Letters}\
  }\textbf {\bibinfo {volume} {97}},\ \bibinfo {pages} {137202} (\bibinfo
  {year} {2006})}\BibitemShut {NoStop}%
\bibitem [{\citenamefont {Compton}\ \emph {et~al.}(2010)\citenamefont
  {Compton}, \citenamefont {Chen},\ and\ \citenamefont
  {Crowell}}]{Compton2010}%
  \BibitemOpen
  \bibfield  {author} {\bibinfo {author} {\bibfnamefont {R.L.}\ \bibnamefont
  {Compton}}, \bibinfo {author} {\bibfnamefont {T.Y.}\ \bibnamefont {Chen}}, \
  and\ \bibinfo {author} {\bibfnamefont {P.A.}\ \bibnamefont {Crowell}},\
  }\bibfield  {title} {\enquote {\bibinfo {title} {Magnetic vortex dynamics in
  the presence of pinning},}\ }\href {\doibase 10.1103/PhysRevB.81.144412}
  {\bibfield  {journal} {\bibinfo  {journal} {Physical Review B}\ }\textbf
  {\bibinfo {volume} {81}},\ \bibinfo {pages} {144412} (\bibinfo {year}
  {2010})}\BibitemShut {NoStop}%
\bibitem [{\citenamefont {Guslienko}\ \emph {et~al.}(2001)\citenamefont
  {Guslienko}, \citenamefont {Novosad}, \citenamefont {Otani}, \citenamefont
  {Shima},\ and\ \citenamefont {Kukamichi}}]{Guslienko2001}%
  \BibitemOpen
  \bibfield  {author} {\bibinfo {author} {\bibfnamefont {K. Y.}~\bibnamefont
  {Guslienko}}, \bibinfo {author} {\bibfnamefont {V.}~\bibnamefont {Novosad}},
  \bibinfo {author} {\bibfnamefont {Y.}~\bibnamefont {Otani}}, \bibinfo
  {author} {\bibfnamefont {H.}~\bibnamefont {Shima}}, \ and\ \bibinfo {author}
  {\bibfnamefont {K.}~\bibnamefont {Fukamichi}},\ }\bibfield  {title} {\enquote
  {\bibinfo {title} {Magnetization reversal due to vortex nucleation,
  displacement, and annihilation in submicron ferromagnetic dot arrays},}\
  }\href {\doibase dx.doi.org/10.1103/PhysRevB.65.024414} {\bibfield  {journal}
  {\bibinfo  {journal} {Physical Review B}\ }\textbf {\bibinfo {volume} {65}},\
  \bibinfo {pages} {024414} (\bibinfo {year} {2001})}\BibitemShut {NoStop}%
\bibitem [{\citenamefont {Burgess}\ \emph {et~al.}(2014)\citenamefont
  {Burgess}, \citenamefont {Losby},\ and\ \citenamefont
  {Freeman}}]{Burgess2014}%
  \BibitemOpen
  \bibfield  {author} {\bibinfo {author} {\bibfnamefont {J.A.J.}\ \bibnamefont
  {Burgess}}, \bibinfo {author} {\bibfnamefont {J.E.}\ \bibnamefont {Losby}}, \
  and\ \bibinfo {author} {\bibfnamefont {M.R.}\ \bibnamefont {Freeman}},\
  }\bibfield  {title} {\enquote {\bibinfo {title} {An analytical model for
  vortex core pinning in a micromagnetic disk},}\ }\href {\doibase
  10.1016/j.jmmm.2014.02.078} {\bibfield  {journal} {\bibinfo  {journal}
  {Journal of Magnetism and Magnetic Materials}\ }\textbf {\bibinfo {volume}
  {361}},\ \bibinfo {pages} {140--149} (\bibinfo {year} {2014})}\BibitemShut
  {NoStop}%
\bibitem [{\citenamefont {Eltschka}\ \emph {et~al.}(2010)\citenamefont
  {Eltschka}, \citenamefont {W{\"o}tzel}, \citenamefont {Rhensius},
  \citenamefont {Krzyk}, \citenamefont {Nowak}, \citenamefont {Kl{\"a}ui},
  \citenamefont {Kasama}, \citenamefont {Dunin-Borkowski}, \citenamefont
  {Heyderman}, \citenamefont {van Driel},\ and\ \citenamefont
  {Duine}}]{Eltschka2010}%
  \BibitemOpen
  \bibfield  {author} {\bibinfo {author} {\bibfnamefont {M.}~\bibnamefont
  {Eltschka}}, \bibinfo {author} {\bibfnamefont {M.}~\bibnamefont
  {W{\"o}tzel}}, \bibinfo {author} {\bibfnamefont {J.}~\bibnamefont
  {Rhensius}}, \bibinfo {author} {\bibfnamefont {S.}~\bibnamefont {Krzyk}},
  \bibinfo {author} {\bibfnamefont {U.}~\bibnamefont {Nowak}}, \bibinfo
  {author} {\bibfnamefont {M.}~\bibnamefont {Kl{\"a}ui}}, \bibinfo {author}
  {\bibfnamefont {T.}~\bibnamefont {Kasama}}, \bibinfo {author} {\bibfnamefont
  {R.E.}\ \bibnamefont {Dunin-Borkowski}}, \bibinfo {author} {\bibfnamefont
  {L.J.}\ \bibnamefont {Heyderman}}, \bibinfo {author} {\bibfnamefont {H.J.}\
  \bibnamefont {van Driel}}, \ and\ \bibinfo {author} {\bibfnamefont {R.A.}\
  \bibnamefont {Duine}},\ }\bibfield  {title} {\enquote {\bibinfo {title}
  {Nonadiabatic spin torque investigated using thermally activated magnetic
  domain wall dynamics},}\ }\href {\doibase 10.1103/PhysRevLett.105.056601}
  {\bibfield  {journal} {\bibinfo  {journal} {Physical Review Letters}\
  }\textbf {\bibinfo {volume} {105}},\ \bibinfo {pages} {056601} (\bibinfo
  {year} {2010})}\BibitemShut {NoStop}%
\bibitem [{\citenamefont {Freeman}\ and\ \citenamefont
  {Hiebert}(2002)}]{Freeman2002}%
  \BibitemOpen
  \bibfield  {author} {\bibinfo {author} {\bibfnamefont {M.R.}\ \bibnamefont
  {Freeman}}\ and\ \bibinfo {author} {\bibfnamefont {W.K.}\ \bibnamefont
  {Hiebert}},\ }\bibfield  {title} {\enquote {\bibinfo {title} {Stroboscopic
  microscopy of magnetic dynamics},}\ }in\ \href@noop {} {\emph {\bibinfo
  {booktitle} {Spin Dynamics in Confined Mangetic Structures I}}},\ \bibinfo
  {editor} {edited by\ \bibinfo {editor} {\bibfnamefont {B.}~\bibnamefont
  {Hillebrands}}\ and\ \bibinfo {editor} {\bibfnamefont {K.}~\bibnamefont
  {Ounadjela}}}\ (\bibinfo  {publisher} {Springer},\ \bibinfo {address}
  {Berlin},\ \bibinfo {year} {2002})\BibitemShut {NoStop}%
\bibitem [{\citenamefont {Badea}\ \emph {et~al.}(2015)\citenamefont {Badea},
  \citenamefont {Frey},\ and\ \citenamefont {Berezovsky}}]{Badea2015}%
  \BibitemOpen
  \bibfield  {author} {\bibinfo {author} {\bibfnamefont {R.}~\bibnamefont
  {Badea}}, \bibinfo {author} {\bibfnamefont {J.~A.}\ \bibnamefont {Frey}}, \
  and\ \bibinfo {author} {\bibfnamefont {J.}~\bibnamefont {Berezovsky}},\
  }\bibfield  {title} {\enquote {\bibinfo {title} {Magneto-optical imaging of
  vortex domain deformation in pinning sites},}\ }\href {\doibase
  10.1016/j.jmmm.2015.01.036} {\bibfield  {journal} {\bibinfo  {journal}
  {Journal of Magnetism and Mangetic Materials}\ }\textbf {\bibinfo {volume}
  {381}},\ \bibinfo {pages} {463--469} (\bibinfo {year} {2015})}\BibitemShut
  {NoStop}%
\end{thebibliography}

%

\end{document}